\documentclass[reprint, prd, twocolumn,floatfix]{revtex4}
\usepackage[english]{babel}
\usepackage[latin1]{inputenc}
\usepackage{enumerate}
\usepackage{amsfonts}
\usepackage{amssymb}
\usepackage{amsmath}
\usepackage{dcolumn}
\usepackage{bm}
\usepackage{color}
\usepackage{graphicx, graphics}
\usepackage{graphicx}

\begin{document}
		\title{Light Pseudoscalar and Axial Spectroscopy using AdS/QCD Modified Soft Wall Model}
		\author{Miguel \'Angel Mart\'{\i}n Contreras}
		\email{miguelangel.martin@uv.cl}
	    \affiliation{Instituto de F\'isica y Astronom\'ia, \\
 Universidad de Valpara\'iso,\\
 Av. Gran Breta\~na 1111, Valpara\'iso, Chile}

	    \author{Santiago Cort\'es}
		\email{jscortesg@unal.edu.co}
		\affiliation{Departamento de F\'{\i}sica, Univ. Nacional de Colombia, 111321 Bogot\'a, Colombia}

		\author{Alfredo Vega}
		\email{alfredo.vega@uv.cl}
	\affiliation{Instituto de F\'isica y Astronom\'ia, \\
 Universidad de Valpara\'iso,\\
 Av. Gran Breta\~na 1111, Valpara\'iso, Chile}
 
		\begin{abstract}
		We describe the mass spectrum of light pseudoscalar and axial mesons in the context of the modified soft wall model with an extra UV cutoff. In order to include the pseudoscalar and axial states, we define an \emph{anomalous dimension} that shifts the conformal dimension of the non-interacting bulk fields such that the parity behavior of those states is included, thus inducing chiral symmetry breaking. This idea contrasts with the usual approach that uses interacting scalar, vector and axial bulk fields to give rise the spectrum. Using the extra UV cutoff approach, we can fit six $\eta$ and six $a_1$ organized in radial trajectories with an RMS error close to 16.9\%. We also confirm that chiral symmetry is restored in this model after checking that highly excited $\rho$ and $a_{1}$ states become degenerate.
		\end{abstract}
	\maketitle
	
\section{Introduction}
\label{intro}
Throughout the last twenty years, the AdS/CFT correspondence \cite{Aharony:1999ti,Maldacena:1997re,Ramallo:2013bua}  has been used to study a wide range of non-perturbative phenomena with significant success. Some examples are given by the low-energy QCD vacuum properties such as confinement, meson spectra and chiral symmetry breaking, along with the description of the quark-gluon plasma state. In order to study all of them, two approaches can be employed: in the first one (top-down), some of these phenomena are modeled via a large $\mathcal{N}=4$ Super Yang Mills theory, which is equivalent to a weak IIB-type supergravity in AdS space. On the other hand, the bottom-up approach starts from a Non IIB-type-SUGRA gravity theory living in an AdS-like background; its dual representation is a Nonconformal QFT with a finite number of $N$ colors, whose behavior is expected to be closer to QCD when $N=3$.

One of the most successful bottom-up approaches is the soft wall model \cite{Karch:2006pv}, whose motivation relies on the fact that radial Regge linear trajectories  are  a consequence of confinement. This feature can be realized if a static dilaton field with a quadratic profile is introduced into the bulk action; this dilaton induces an energy scale $c$ that defines Regge trajectories so that the squared mass of the hadrons scales as $M^2=A \,c^2+B$, where $A$ and $B$  are model-dependent parameters \cite{Vega:2008te}. This bottom-up approach was used to study glueballs \cite{BoschiFilho:2002ta,BoschiFilho:2012xr,Capossoli:2015ywa,Capossoli:2016kcr}, hadron form factors \cite{Gutsche:2015xva} and light meson spectra \cite{Colangelo:2008us}.

Despite the success of the soft wall model, it cannot address some other QCD aspects such as chiral symmetry breaking, heavy quarkonium spectra, mesonic decay constants or thermal and density properties of a colored medium. These questions motivated improvements to the original soft wall model, for instance, its dynamical version \cite{Gherghetta:2009ac,Li:2012ay}, as well as the introduction of an extra UV cutoff \cite{Braga:2015jca,Braga:2016wkm,Braga:2017bml}. The latter approach allows to include an extra energy scale given by the locus $z_0$ of a D-brane. Other interesting approaches to Regge trajectories in holographic models are summarized in \cite{Huang:2007fv,Fang:2016uer}; one of them \cite{Fang:2016uer}, also known as the IR--improved soft wall model, constructs the meson mass spectrum for chiral partners considering a set of scalar, axial and vector fields that interact within the bulk. 

Our main purpose is to analyze the application of the modified soft wall model to describe light pseudoscalar and axial mesonic states. Thus, we organize the present paper as follows: in section \ref{sec:Holset} we show the holographic setup for a modified soft wall model with an extra UV cutoff defined as a D-brane located at $z_0$, where mesons are dual to free bulk fields. We focus section \ref{sec:hdron} to discuss how hadrons are identified in AdS/QCD models and show how to include $\eta$ and $a_1$ mesons using a \emph{anomalous dimension}  added to the conformal dimension; the latter is utilized to define the bulk mass for each bulk field, hence allowing to put the hadronic identity of the mesons at hand by introducing an extra parameter $\Delta_P$. 
In section \ref{holo-des} and \ref{holo-a1} we construct the bulk-to-boundary propagators, the 2-point function and the mass spectrum for both the $\eta$ and the $a_1$ mesons using the extra UV cutoff concept developed in \cite{Braga:2015jca,Cortes:2017lgz}. Numerical results are presented in section \ref{numerical} and finally, some conclusions are given in section \ref{conclusions}. 

\section{Holographic Setup}
\label{sec:Holset}
Our starting point is the model introduced in \cite{Braga:2015jca} adapted to the light scalar ($f_0$) and vector ($\rho$) radial trajectories.  
We thus define the usual AdS Poincare patch with an extra UV cutoff as

\begin{eqnarray}
dS^2&=&g_{MN}\,dx^M\,dx^N\\ \notag
&=&\frac{R^2}{z^2}\left[dz^2+\eta_{\mu \nu}\,dx^\mu\,dx^\nu\right]\,\Theta\left(z-z_0\right),	
\end{eqnarray}

where $\Theta\left(x\right)$ is the Heaviside step function, $R$ is the AdS radius, and $z_0$ is the UV cut-off defined by the locus of a D-brane. The associated action reads

\begin{equation}
I=I_\text{Scalar}+I_\text{Vector},
\end{equation}

with

\begin{multline}\label{scalars}
 I_\text{Scalar}=-\frac{1}{2\,g_S^2}\int d^5\,x\,\sqrt{-g}\,e^{-\Phi\left(z\right)}\left[g^{MN}\,\partial_M\,S\,\partial_N\,S\right.\\
 \left.+M_5^2\,S^2\right],
\end{multline}

\begin{multline}\label{vectors}
 I_\text{Vector}=-\frac{1}{2\,g_V^2}\int d^5\,x\,\sqrt{-g}\,e^{-\Phi\left(z\right)}\left[\frac{1}{2}F_{MN}\,F^{MN} \right.\\
 \left.+\tilde{M}_5^2\,g^{MN}\,A_M\,A_N\right].   
\end{multline}

$F_{MN}=\partial_M\,A_N-\partial_N\,A_M$ is the field strength related to the U(1) field $A_M\left(z,x^\mu\right)$, the coupling $g_{S\,(V)}$ is a constant that adjusts units on the scalar (vector) sector, and $M_5$ ($\tilde{M}_5$) is the bulk mass that fixes the hadronic identity for scalar (vector) states. The term $\Phi\left(z\right)=\kappa^2\,z^2$ is the static quadratic dilaton profile used in the original soft wall model.

From these actions, we obtain the equations of motion for the bulk fields. Then, after imposing Dirichlet boundary conditions at $z_0$, we construct the bulk-to-boundary propagators that transport the holographic information of the fields to the D-brane at $z_0$. Finally, by derivating twice the bulk-to-boundary propagator with respect to the sources, which can be understood as a holographic equivalent to the 2-point function,  we extract the mesonic mass spectra via their pole expansions. 

\section{Hadronic identity in the AdS/QCD models}\label{sec:hdron}

In this section we address how to describe hadrons in AdS/QCD models. For example, if we consider scalar bulk fields according to the holographic dictionary, their respective non-perturbative  solutions of the bulk equations of motion scale as $z^{\Delta-4}$ in the $z\to 0$ limit; furthermore, they are dual to some $\mathcal{O}$ CFT operator whose 2-point function is found to be \cite{Witten:1998qj} 
\begin{equation*}
\langle\mathcal{O}(x)\,\mathcal{O}(0)\rangle \propto \frac{1}{\left|x\right|^{2\,\Delta}},
\end{equation*}

which implies that $\Delta$ is the \emph{scaling} dimension of the operator $\mathcal{O}$. This can be extrapolated to higher spin  fields, thus obtaining a similar behavior. Hence, fixing $\Delta$ gives the scaling properties of $\mathcal{O}$ \footnote{This is the key point to introduce hadrons in AdS/QCD.}.  

The latter $\Delta$ term is also related to the operators that create hadrons, as well as to their
respective hadronic identity. In order to clarify this, we can analyze the case of mesons: it is known that they are created by $q\bar{q}$ operators whose dimension is three, therefore, taking $\Delta=3$ aids us to holographycally identify them as mesonic states. 

Bulk mass terms in the action affect the low $z$ limit of the solutions, therefore, $\Delta$ is to be connected to $M_5^2$. In general, for a field of spin $s$ we have  \cite{Aharony:1999ti,Witten:1998qj}:
\begin{equation}
    2\,\Delta_\pm=4\pm\sqrt{\left(4-2\,s\right)^2+4\,M_5^2\,R^2}.
\end{equation}

As a consequence, the \emph{hadronic} dimension is related with the bulk mass as 

\begin{equation}
    M_5^2\,R^2=\Delta\,\left(\Delta-4\right)-s\left(s-4\right).
\end{equation}
 
The latter equation defines how we can introduce mesons of any spin into the theory via the bulk mass $M_5^2$. As an illustration, for \emph{scalar mesons} we have $M_5^2\,R^2=-3$, whereas for \emph{vector mesons}, $M_5^2\,R^2=0$. Notice that we do not distinguish between $M_5^2$ and $\tilde{M}_5^2$ since we give the corresponding values for each case.   

This bulk mass definition is solely associated to s-wave ($L=0$) mesonic states with isospin ($I$) fixed to zero ($f_0$ mesons \cite{Colangelo:2008us}) and one ($\rho$ mesons \cite{Karch:2006pv}). However, if we want to include other states with $L$ different from zero, we can introduce some sort of  \emph{twist} operator $\tau$ to the conformal dimension $\Delta\rightarrow \Delta+\tau$ in order to raise the value of $L$, as it was exposed in \cite{Vega:2008te,BoschiFilho:2012xr,Vega:2010ne,Forkel:2010gu,Gutsche:2011vb,Branz:2010ub}). The inclusion of isospin in this sort of AdS/QCD models is still at discussion. 

The idea of an anomalous dimension can be used to explore other properties such as chiral symmetry breaking ($\chi$SB) and its relation with vector-axial meson spectra \cite{Huang:2007fv}. As it is well known, these particles become non degenerate after $\chi$SB, hence, they got to be distinguishable. In order to attain this, we introduce the anomalous dimension $\Delta_P$ to identify the parity of the mesonic state considered at hand, thus inducing a mechanism to break chiral symmetry by modifying the bulk mass as follows:

\begin{align} \label{mass-anomalous}	
M_5^2\,R^2&=\left(\Delta_\text{Phys}+\Delta_P\right)\left(\Delta_\text{Phys}+\Delta_P-4\right) \notag \\ 
&-s\left(s-4\right).
\end{align}

It must be kept in mind that the values for $\Delta_P$ are constrained since the bulk masses have to be bounded by below due to the stability of the solutions, according to the Breitenholner--Freedman limit \cite{Breitenlohner:1982bm,Breitenlohner:1982jf,Moroz:2009kv}. The choices of $\Delta_P$ are exposed in table \ref{tab:summaryM5} for each mesonic field. 

\begin{table}
\caption{This table summarizes the fixing of $\Delta_P$ and the value of $M_5^2\,R^2$ on each case of interest. Notice that pseudoscalar and axial cases are allowed by the stability conditions given in \cite{Breitenlohner:1982bm,Breitenlohner:1982jf,Moroz:2009kv,Case:1950an}. Scalar and vector cases were discussed in \cite{Cortes:2017lgz}.}
\centering 
\begin{tabular}[c]{c||c||c|}
\textbf{Meson Identity}  & $\Delta_P$ & $M_5^2\,R^2$\\
\hline
\hline
Scalar meson       & 0    & $-3$\\
\hline
Vector meson       & 0    & $0$\\
\hline
Pseudoscalar meson & $-1$ & $-4$\\
\hline 
Axial vector meson & $-1$ & $-1$\\
\hline
\end{tabular}

\label{tab:summaryM5}
\end{table}

\section{Holographic description of pseudoscalar mesons}\label{holo-des}
Using the action (\ref{scalars}), we describe the mass spectrum of the $\eta$ mesons trajectory. At first glance, there is no difference in the scalar fields used to describe scalar or pseudoscalar mesons since the bulk mass is the same for both of them. This motivates us to include the $\Delta_P$ anomalous dimension, as we suggest in equation (\ref{mass-anomalous}). The connection with parity becomes clear when we check how $f_0$ ($J^{PC}=0^{++}$) and $\eta$ ($J^{PC}=0^{-+}$) mesons behave with it: the $f_0$ trajectory is invariant under parity transformations whereas the $\eta$ one is  not. This hints the possible (phenomenological) connection with the 
anomalous dimension \footnote{This is why we call it $\Delta_{P}$.}. 

The $f_0$ trajectory is constructed with good precision by considering $M_5^2\,R^2=-3$ (See \cite{Cortes:2017lgz}). Hence, we can fix $\Delta_P=0$ for the $f_0$ family. From the Breitenholner-Freeman limit, we can infer that the possible value of $\Delta_P$ for $\eta$ mesons is $\Delta_P=-1$; then, for this sort of background, we should have  $M_5^2\,R^2\geq-4$ for pseudoscalar fields in order to generate stable solutions \cite{Breitenlohner:1982bm,Breitenlohner:1982jf,Moroz:2009kv}. This procedure is similar to that showed in \cite{Huang:2007fv}, where the bulk mass of the axial fields is shifted from its initial null value so that their Regge trajectory is to be splitted from the one associated to the vector fields. Thus, an auxiliar scalar field does not need to be considered to break chiral symmetry, as happens in \cite{Li:2013oda}, where the splitting between vector and axial mesons is produced by an interaction term between the auxiliar and axial fields in the associated quadratic action of these last ones.

The equation of motion for the pseudoscalar mesons is obtained by doing variations on the action (\ref{scalars}). After Fourier transforming and imposing the on-shell mass condition $q^2=m_n^2$, we arrive to the following expression:

\begin{equation}\label{eqnmotion}
	\partial_z\left[\frac{e^{-\kappa^2\,z^2}}{z^3}\,\partial_z\,S\right]+	\frac{e^{-\kappa^2\,z^2}}{z^3}\,q^2\,S+\frac{4\,e^{-\Phi}}{z^5}S=0.
\end{equation}

According to the well-established holographic recipe for hadrons \cite{Karch:2006pv,Braga:2015jca,Branz:2010ub,Vega:2009zb}, it is customary to obtain the mass spectrum from the pole expansion of the 2-point function constructed with the solutions of eq. (\ref{eqnmotion}). We will focus on this method later. 

\subsection{Pseudoscalar bulk to boundary propagator}
As the holographic prescription given by \cite{Witten:1998qj} dictates, the non-normalizable modes generate the QFT operators when they are evaluated at the conformal boundary. In our case, we will put the boundary at $z_0$; after considering Dirichlet conditions for it, we will find that those modes generate the desired light pseudoscalar meson spectrum.

Let us take eq. (\ref{eqnmotion}) to perform the following transformation in the pseudoscalar field: $S(z,q)=S_0(q)\,\mathcal{V}(z,q)$. The field $\mathcal{V}(z,q)$ is the bulk-to-boundary propagator that projects holographic information onto the boundary related with the source operator $S_0(q)$. The bulk-to-boundary propagator holds with the boundary condition $\mathcal{V}(z_0,q)=1$. After considering this, the equation of motion now reads

\begin{equation}\label{eqnmotionV}
	\partial_z\left[\frac{e^{-\kappa^2\,z^2}}{z^3}\,\partial_z\,\mathcal{V}\right]+\frac{e^{-\kappa^2\,z^2}}{z^3}\,q^2\,\mathcal{V}+\frac{4\,e^{-\kappa^2\,z^2}}{z^5}\mathcal{V}=0.
\end{equation}

The solution for this equation is written in terms of Kummer confluent hypergeometric $_1F_1$ functions as follows:

\begin{equation}\label{etaeq}
    \mathcal{V}_\eta\left(z,q\right)=\frac{z^2\,_1F_1\left(1-\frac{q^2}{4\kappa^2},\,1,\,\kappa^2\,z^2\right)}{z_0^2\,_1F_1\left(1-\frac{q^2}{4\kappa^2},\,1,\,\kappa^2\,z_0^2\right)}.
\end{equation}
We used the subindex $\eta$ in the latter equation to emphasize that this solution will bring the mass spectrum of $\eta$ mesons. 

\subsection{Pseudoscalar 2-point function}
\label{sec:pseudo2p}

In order to obtain the mass spectrum, we need to construct the corresponding 2-point function so its pole expansion is to be found. Holographically speaking, the 2-point function is constructed by applying the saddle point approximation on the on-shell boundary action via its second derivative. 

The on-shell boundary action in the scalar case is obtained by evaluating the solution (\ref{etaeq}) in the action (\ref{scalars})

\begin{multline}
  I^\text{On-shell}_{\text{Bndary, }\eta} = \frac{R^3}{g^2_S}\,\int \frac{d^4\,q}{\left(2\pi\right)^4}\,\frac{e^{-\kappa^2\,z^2}}{z^3}\,S_0\left(q\right)\,S_0\left(-q\right)\\
  \left.\times \mathcal{V}\left(z,q\right)\partial_z\,\mathcal{V}\left(z,-q\right)\right|_{z=z_0},  
\end{multline}

where we have used the Gauss theorem on the D-brane at $z_0$ along the normal unitary vector $\hat{n}_z=-\frac{1}{\sqrt{g^{zz}}}\,$.  

After taking the second derivative with respect to the sources $S_0$,  we arrive  to the following expression for the 2-point function $\Pi(q^2)$ \cite{Braga:2015jca}:

\begin{equation*}
\Pi_\eta\left(q^2\right)=\frac{\delta^2\, I^\text{On-shell}_{\text{Bndary, }\eta}}{\delta\,S_0\left(-q\right)\,\delta\,S_0\left(q\right)}=\frac{R^3}{g^2_S}\,\left. \frac{e^{-\kappa^2\,z^2}}{z^3}\,\partial_z\,\mathcal{V}\left(z,q\right)\right|_{z_0}.    
\end{equation*}
 
 After evaluating the bulk-to-boundary propagator (\ref{etaeq}) in the expression above we obtain that

\begin{multline}
\Pi_\eta\left(q^2\right)=-\frac{R^3}{g_S^2}\frac{e^{-\kappa^2\,z^2_0}}{z_0^3}\left[\frac{2}{z_0}+\right.\\
\left.\left(1-\frac{q^2}{4\,\kappa^2}\right)\frac{2\,\kappa^2\,z_0\,_1F_1\left(2-\frac{q^2}{4\,\kappa^2},\,2,\kappa^2\,z^2_0\right)}{_1F_1\left(1-\frac{q^2}{4\,\kappa^2},\,1,\kappa^2\,z^2_0\right)}\right]\label{eta-2f}.     
\end{multline}    
    
The mass spectrum can be read from the poles of (\ref{eta-2f}), which are given by the roots $\chi_n$ of the Kummer hypergeometric function $_1F_1(a\, ,b\, ,x)$ in the the denominator as \cite{Cortes:2017lgz} 

\begin{equation}
    _1F_1\left(1-\chi_n,\,1,\kappa^2\,z^2_0\right)=0,
\end{equation}

where $\chi_n=M_n^2/4\,\kappa^2$ is the root spectrum and $M_n^2=q^2$ are the masses of the pseudoscalar mesons. Thus, the mass spectrum is given by

\begin{equation}
M_n^2=4\,\kappa^2\,\chi_n\left(\kappa,\,z_0,\Delta_P\right).	
\label{eqn:scalarmass2}
\end{equation}
 
The result above assures that the mass spectrum   is a linear radial Regge trajectory defined by the parameters $z_0$, $\kappa$ and $\Delta_P$. In general, the roots of the Kummer confluent hypergeometric  function increase with the radial excitation number $n$, as well as the masses (as we expected). The respective numerical results are presented in table \ref{tab:pseudomes}. Notice that if we fix $\Delta_P=0$, we obtain $M_5^2\,R^2=-3$, thus giving the light scalar meson spectrum discussed in \cite{Cortes:2017lgz}.

\section{Holographic description of axial mesons}\label{holo-a1}
The case of axial mesons is not so different from the one discussed in the latter section; in fact, their equations of motion exhibit a similar form compared to the pseudoscalar version. We confirm this after obtaining the following set of equations of motion for the vector field $A_m$ from the action (\ref{vectors}) :

\begin{multline}
   \frac{1}{\sqrt{-g}}\,\partial_M\,\left[\sqrt{-g}\,e^{-\kappa^2\,z^2}\,g^{MR}\,g^{NP}\,F_{RP}\right]+\\
   \,e^{-\kappa^2\,z^2}\,g^{MN}\,A_N=0.  
\end{multline}

It is well known from particle physics that the $a_1$'s ($J^{PC}=1^{++}$) have different parity compared to the light $\rho$ vector states ($J^{PC}=1^{--}$). As in the pseudoscalar case, this also suggest us to modify the conformal dimension on the bulk vector fields in order to mimic this difference. By looking closely to eq. (\ref{mass-anomalous}), we observe that for mesons with $\Delta=3$, $M_5^2\,R^2=0$ holds, thus ensuring the gauge invariance. This case was discussed in \cite{Cortes:2017lgz} in the context of $\rho$ mesons. 

Following the procedure taken for the pseudoscalar case, we introduce the \emph{anomalous dimension} related to parity $\Delta_P$. In the vector case this implies that the bulk mass is non-zero, so we need to review the gauge structure and the stability of the solutions. 

The former issue can be addressed as follows: for the $\rho$ Regge trajectory with $M_5^2\,R^2=0$, the equations of motion are gauge invariant and we can fix the gauge $A_z=0$. When considering axial mesons, the associated field is massive, hence, gauge invariance is broken. Nevertheless, following the ideas exposed in \cite{Braga:2015lck}, the gauge $A_z=0$ also holds in this case: the equation of motion for the $A_z$ component reads $\Box \,A_z-\partial_z\,\left(\partial_\mu\,A^\mu\right)-M_5^2\,A_z=0$. Fixing $A_z=0$ allows us to impose a plane wave expansion in the solutions since $\partial_\mu\,A^\mu=0$ is still valid. 

The latter issue is resolved by considering the lower value of the bulk mass that grants us to have stable solutions, as in the scalar case. Consequently, we find that stable solutions should have $M_5^2\,R^2 \geq -1$. Therefore, we will fix $\Delta_P=-1$, as in the $\eta$ meson case.

Following these ideas, and imposing the condition $A_z=0$, we obtain (after Fourier transforms) the equation of motion for the axial transverse modes $A_m\left(z,q\right)$:

\begin{equation}
\partial_{z}\left[\frac{e^{-\kappa^2\,z^2}}{z}\,\partial_z\,A_m\right]+q^2\,\frac{e^{-\kappa^2\,z^2}}{z}\,A_m+\frac{e^{-\kappa^2\,z^2}}{z^{3}}A_m=0.
\label{eqn:eomaxial0}
\end{equation}

As we did in the pseudoscalar case, we will generate the bulk-to-boundary propagator $\mathcal{V}(z,q)$ from this equation and use it to construct the 2-point function in such a way that the mass spectrum is to be spawned. 

\subsection{Axial bulk-to-boundary operator}

We need to construct the bulk-to-boundary propagator before describing the respective axial 2-point function. To do so, we will follow a similar path as in the pseudoscalar case. First of all, we define the transverse vector field as $A_m(z,q)=A_m^0(q)\,\mathcal{V}(z,q)$ and impose the Dirichlet boundary condition $\mathcal{V}\left(z_0,q\right)=1$.  The equation for the bulk to boundary propagator now reads

\begin{equation}
\partial_{z}\left[\frac{e^{-\kappa^2\,z^2}}{z}\,\partial_z\,\mathcal{V}\right]+q^2\,\frac{e^{-\kappa^2\,z^2}}{z}\,\mathcal{V}+\frac{e^{-\kappa^2\,z^2}}{z^{3}}\mathcal{V}=0,
\label{eqn:eomaxial}
\end{equation}
whose solution is given in terms of Kummer confluent hypergeometric functions $_1F_1$ as 

\begin{equation}\label{bulk-to-boundary-V}
    \mathcal{V}_{a_1}\left(z,q\right)=\frac{z\,_1F_1\left(\frac{1}{2}-\frac{q^2}{4\kappa^2},\,1,\,\kappa^2\,z^2\right)}{z_0\,_1F_1\left(\frac{1}{2}-\frac{q^2}{4\kappa^2},\,1,\,\kappa^2\,z_0^2\right)}.
\end{equation}

We have used the subindex $a_1$ just to emphasize that we are dealing with the $a_1$ meson trajectory. We can focus now on the axial boundary action and the 2-point function. 

\subsection{Axial 2-point function}
 As we showed in \ref{sec:pseudo2p}, we evaluate the action (\ref{vectors}), integrate over the $z$-direction and obtain the boundary action for the vector case as it is showed below:

\begin{multline}
I_{\text{Bndry, } a_1}^\text{On-shell}=\frac{R}{g_V^2}\int\frac{d^4\,q}{\left(2\,\pi\right)^4}\frac{e^{-\kappa^2\,z^2}}{z}\eta^{\mu\nu}\,A_\mu^0\left(q\right)\,A_\nu^0\left(-q\right)\\
\times\left.\partial_z\,\mathcal{V}\left(z,q\right)\right|_{z_0}.
\end{multline}

In the latter equation, $\eta_{\mu \nu}$ is the Minkowski metric tensor. Applying the holographic prescription, we now take the second derivative of the on-shell boundary action with respect to the sources $A_\mu^0\left(q\right)$ to obtain the 2-point function as

\begin{eqnarray*}
\Pi^{\mu \nu}_{a_1}\left(q^2\right)&=&\frac{\delta^2\,I^\text{On-shell}}{\delta A_\mu^0\left(q\right)\,\delta A_\nu^0\left(-q\right)}\\
&=&\frac{R}{g_v^2}\,\eta^{\mu \nu}\left.\frac{e^{-\kappa^2\,z^2}}{z}\,\partial_z\,\mathcal{V}\left(z,q\right)\right|_{z_0}.
\end{eqnarray*}

Evaluating the solutions for the bulk to boundary propagator expressed in \eqref{bulk-to-boundary-V}, we arrive to the following axial vector 2-point function:

\begin{multline}
  \Pi^{\mu \nu}_{a_1}\left(q^2\right)=   -\frac{R}{g_V^{2}}\frac{e^{-\kappa^2\,z_0^2}}{z_0}\left[\frac{1}{z_0}+\right.\\
  \left.\left(\frac{1}{2}-\frac{q^2}{4\kappa^2}\right)\frac{2\,\kappa^2\,z_0\,_1F_1\left(\frac{3}{2}-\frac{q^2}{4\kappa^2},2,\kappa^2\,z_0^2\right)}{_1F_1\left(\frac{1}{2}-\frac{q^2}{4\kappa^2},1,\kappa^2z_0^2\right)}\right]\,\\
  \times\left(\eta^{\mu \nu}-\frac{q^\mu\,q^\nu}{q^2}\right). \label{eqn:axialp}
\end{multline}

The latter result grants us to obtain a pole expansion that depends on the zeros of the denominator of (\ref{eqn:axialp}), which is defined in terms of the roots of the Kummer confluent hypergeometric function as

\begin{equation}
    _1F_1\left(\frac{1}{2}-\xi_n,1,\kappa^2\,z_0\right)=0,
\end{equation}

where $\xi_n=M_n^2/4\kappa^2$ is the root spectrum and $M^2_n=q^2$ defines the mass of the $a_1$ mesons. Therefore, the mass spectrum for the axial mesons is 

\begin{equation}\label{axial-mass}
    M_n^2=4\,\kappa^2\,\xi_n\left(\kappa, z_0, \Delta_P\right).
\end{equation}

We have put again the explicit dependence of the mass spectrum with the parameters $\kappa$, $z_0$ and $\Delta_P$. As showed in the pseudoscalar case (\ref{sec:pseudo2p}), the roots of the Kummer confluent hypergeometric functions are increasing with the radial excitation number $n$, implying that the masses are increasing with $n$ as we expected. Numerical results are presented in table \ref{tab:axialmes} for this specific case.

\section{Numerical Results}\label{numerical}

As we mentioned above, this work follows the ideas exposed in \cite{Braga:2015jca,Cortes:2017lgz}. Hence, in order to have universality in these approaches, we should define in a similar way the choosing of parameters $\kappa$ and $z_0$: in those works, it was discussed that $\kappa$ should be \emph{flavor dependent} since it is related to the $c$ and $b$ quark content inside the mesons. Thanks to this, the meson spectra of heavy quarkonia were well described by picking proper values that minimized the errors produced after comparing the results with experimental data. It is worth to notice that the ground states of these heavy mesons have masses whose corresponding values are remarkably close to the algebraic sum of the current  quark masses.

The set of light unflavored mesons described in the present manuscript is entirely made of unflavoured $u$, $d$ and $s$ bound states. However, their masses are not given wholly by the algebraic sum of $u,\,d\text{ and }s$ current masses; as a matter of fact, light mesons have masses of hundreds of MeVs as a consequence of the binding energy between these quarks. Because of this, and reminding that these fermion particles can be described in terms of an approximate $SU(3)$ symmetry, we choose $\kappa=0.45$ GeV as it was fixed in \cite{Cortes:2017lgz}; furthermore, it is an energy scale whose value resembles that of the constituent masses of these light quarks. 

On the other hand, $z_0$ was associated to the natureness of the strong interaction that bounds the pair of quarks into the meson. Heavy quarkonia states were well described by choosing $z_{0}=4\text{ GeV}^{-1}$ \cite{Braga:2015jca}. Thus, following \cite{Cortes:2017lgz}, and considering that we are dealing with unflavored $q\overline{q}$ mesons, we will keep $z_0=5$ GeV$^{-1}$.

Since binding energy is larger between light flavors, it is expected $z_{0}$ to be changed. Nevertheless, this change has to be made so that most of the mesons considered are properly predicted by the model.

 Notice that when comparing with the soft wall model \cite{Karch:2006pv}, the dilaton parameter $\kappa$ has a different meaning: in the former, it is connected with the Regge slope, whilst in this approach is related to the constituent quark mass. \color{black}{The Regge parameter in our case comes from the combination of both $z_0$ and $\kappa$.} 

\subsection{$\eta$ mesons}
We present in table \ref{tab:pseudomes}  the numerical results for $\eta$ mesons obtained with the holographic description given above compared to the experimental data. It is important to notice that we take the first state in the radial trajectory as the $\eta$ instead of the $\eta^{'}$ meson since we want to describe the $I=0$ pseudoscalar state of the $SU(3)$ meson octet.  

If we compare our first state of the fitted trajectory to the $\eta^{'}$  \cite{Anisovich:2000kxa,Ebert:2009ub}, we obtain a very good correspondence since its associated error is close to 1.77 $\%$. It is appropriate to recall that the $\eta'$ state appears as a singlet state when we consider the meson spectra built from the fundamental and anti-fundamental representations of $SU(3)$; moreover, its mass difference can be described by effects of the chiral anomaly \cite{Witten:1979vv}. However, we do not consider chiral effects directly in this approach since our goal is to fit the $\eta$ mass spectrum using only the modified soft wall parameters $\kappa$ and $z_{0}$, along with the anomalous dimension $\Delta_{p}$. Chiral approaches on the AdS/QCD models are discussed in previous works such as \cite{Colangelo:2008us,Vega:2010ne,Gutsche:2012ez}.

\begin{table}[tbh]
\caption{Mass spectrum for $\eta$ pseudoscalar mesons with $\kappa=0.45$ GeV and $z_0=5.0$ GeV$^{-1}$. Experimental values are obtained from \cite{Tanabashi:2018oca}. For the $\eta(1760)$ and $\eta(2225)$ states, their masses are taken from \cite{Wang:2017iai,Li:2008mza}.}
\centering
\begin{tabular}[c]{|c|c|c|c|c|}
\hline
\hline
\multicolumn{5}{|c|}{$\eta$ trajectory with $\Delta_P=-1$}
\\
\hline
\hline
$n$ & State & $M_\text{Exp}$ (MeV) & $M_\text{Th}$ (MeV) & $\%M$ \\
\hline
\hline
1 &$\eta(550)$ & $547.86\pm 0.017$ & 975.25  & 43.8 \\ 
2& $\eta(1295)$& $1294\pm 4$       & 1233.6  & 4.9   \\
3& $\eta(1405)$& $1408.8\pm 1.8 $  & 1455.3  & 3.2  \\
4& $\eta(1475)$& $1476\pm 4$       & 1652.9  & 10.7 \\ 
5& $\eta(1760)$& $1760\pm 11$      & 1829.2  & 3.8  \\
6& $\eta(2225)$& $2216\pm 21$      & 1992.7  & 11.3 \\
\hline
\hline 
\end{tabular}
\label{tab:pseudomes}
\end{table}

\subsection{$a_1$ mesons}
Our predictions for the masses of three PDG-listed axial mesons are showed in table \ref{tab:axialmes}. It is worth to notice that the first excitation, i.e., the one with a mass of 809.0 MeV, is not heavy enough to be matched with the $a_1(1260)$. In fact, the $a_1$ meson is an orbital excitation (and also an axial partner) of the $\rho$ meson in the constituent quark model. Besides, the axial meson has a dominant decay given by $a_1\rightarrow \pi\,\rho$; in other words, the phase space of the first state is not adequately large to contain this process since it is required to have a minimum energy around of 915 MeV. 

These observations could lead to consider that the ground state in the model could be missed. But, looking closely to other AdS/QCD models with static quadratic dilaton as \cite{Karch:2006pv} or \cite{Ballon-Bayona:2020qpq} (see table 3), the holographic ground state is also far from the experimental one. This fact seem to be customary of these sorts of models.

To avoid this issue there are two paths: first to consider that dynamical holographic effects due to the \emph{vacuum expectation value} (vev) encoded into the scalar bulk field, that couples with the axial vector bulk field, as it was done in \cite{Gherghetta:2009ac} and \cite{Li:2013oda} in the dynamical background context. The other possibility is to consider self-interactions of the bulk scalar field along with an analytic extension of the bulk mass. As a matter of fact, in all of these AdS/QCD model, the mesonic spectra come from  a holographic potential that the following general form

\begin{equation}\label{holo-pot}
V(z)=A\,z^2+\frac{B}{z^2}+C,    
\end{equation}

\noindent where the constants $A$, $B$ and $C$ are determined by the model parameters at hand. In particular, the bulk mass contribution is contained in $1/z^2$ term, i.e., the $B$ constant. Therefore, if we modify $M_5^2\,R^2\rightarrow f(z)$, with the right asymptotics in order to match any of the general terms in the holographic potential, we can modify the meson spectra. This was done in  \cite{Vega:2010ne,Vega:2011tg,Fang:2016nfj,Ballon-Bayona:2020qpq}. 

The idea exposed here of modifying the conformal dimension by the introduction of an anomalous one can be classified in this sort of models, since we modify the bulk mass also, but not continously.


\color{black}{In order to be consistent, we include other possible $a_1$ meson candidates that have been excluded from PDG and other recent findings since a confirmation is needed. Information about this can be found in \cite{Chen:2015iqa}.}


Another fact interesting to see is that these soft wall-like models are expected to have linear spectrum for high values of $n$ due to the presence of the dilaton field $\Phi(z)=\kappa^2\,z^2$, so the lower states are strongly affected by the columbian part of the holographic potential that can be set from the Sturm-Liouville equations for $\eta$  or $a_1$ mesons, summarized in the expressions (\ref{eqnmotion}) and (\ref{eqn:eomaxial}), respectively\footnote{This potential is constructed using the Bogoliubov transformation $S(q,z)=e^{B(z)/2}\,\phi(z)$, with $B(z)=\kappa^2\,z^2+\log\,z$ in the scalar case and $\phi(z)$ taken as a normalizable solution. See \cite{Karch:2006pv} for details.}. 

However, despite this fact, we could fit six states (three well-established states at PDG, along with three candidates suggested in \cite{Chen:2015iqa,Anisovich:2001pn,Nozar:2008aa}), with an RMS error for these six states fitted with three parameters near to 11\%. This first non-physical holographic state could be suppressed using isospectral techniques \cite{Vega:2016gip} that allow us to extract the first state in the spectrum.  

\begin{table}[tbh]
\caption{Mass spectrum for $a_{1}$ axial mesons with $\kappa=0.45$ GeV and $z_0=5.0$ GeV$^{-1}$. Experimental values are obtained from \cite{Tanabashi:2018oca}. For the $a_{1}(1420)$ state, its mass is read from \cite{Adolph:2015pws}, whereas the masses of $a_1(1930)$, $a_1(2095)$ and $a_1(2270)$ are read from \cite{Anisovich:2001pn}. }
\centering
\begin{tabular}[c]{|c|c|c|c|c|}
\hline
\hline
\multicolumn{5}{|c|}{$a_1$ trajectory with $\Delta_P=-1$}\\
\hline
\hline
$n$ & State & $M_\text{Exp}$ (MeV) & $M_\text{Th}$ (MeV) & $\%M$ \\
\hline
\hline
1 & $a_{1}(1260)$ & $1230\pm 40$             & $808.1$ & $52.2$   \\ 
2 & $a_{1}(1420)$ & $1414^{+ 15}_{- 13}$ & $1114.7$ & $26.8$ \\
3 & $a_{1}(1640)$ & $1654\pm 19 $            & $1351.3$ & $22.4$ \\
4 & $a_{1}(1930)$ & $1930^{+ 19}_{- 70}$            & $1558.7$ & $23.8$ \\
5 & $a_{1}(2095)$ & $2096\pm 17 \pm 121$            & $1744.3$ & $20.1$ \\
6 & $a_{1}(2270)$ & $2270^{+ 55}_{- 40}$         & $1913.4$ & $18.6$ \\
\hline
\hline 
\end{tabular}
\label{tab:axialmes}
\end{table}

Following \cite{Braga:2015jca}, we can test the predictability of the model developed here with the RMS error for estimating $N$ resonances using $N_p$ parameters as

\begin{equation}
    \delta_\text{RMS}=\sqrt{\frac{1}{N-N_p}\,\sum_i^N\,\left(\frac{\delta\mathcal{O}_i}{\mathcal{O}_i}\right)^{2}},
\end{equation}

where $\mathcal{O}_i$ is the experimental mean value of a given operator with an absolute uncertainty $\delta \mathcal{O}_i$. 

In our case, this extra UV cutoff model can fit 26 states, which can be distinguished by six pseudoscalar, six axial, eight scalar and six vector mesons (we take here our previous results given in \cite{Cortes:2017lgz}). Therefore, after considering three parameters given by $\kappa$, $z_{0}$ and $\Delta_{P}$, the value we obtain for $\delta_{\text{RMS}}$ is close to $23.6$ $\%$.

There is something we want to point out, and it is related to the fact that chiral symmetry is restored in highly excited states since the mass multiplets of the axial and vector mesons, also called chiral partners, become degenerated; this is achieved after checking the value of the mass difference $m_{a_{1}}^{\,2}-m_{\rho_{1}}^{\,2}$, where we take into account the respective results given for the highest $\rho$ and $a_{1}$ mesons in \cite{Cortes:2017lgz} and table \ref{tab:axialmes}, i.e., 2.142 GeV and 2.215 GeV. We thus obtain that this difference is approximately 0.318 GeV$^{2}$, a value that is in good accordance with the one previously mentioned in \cite{Huang:2007fv}. 

\section{Conclusions}\label{conclusions}
In this work we have used the modified soft wall model with an extra UV cutoff, an approach that has proven to be a good approximation to study mass spectra of heavy quarkonium \cite{Braga:2015jca} and in the lightest scalar and vector meson sector \cite{Cortes:2017lgz}. The model has three parameters $\kappa$ related to the constituent quark mass, $z_0$ connected with the strong interaction \emph{natureness} and $\Delta_P$ that takes account for the parity change in the analized spectra. This $\Delta_P$ parameter is strongly tied by the stability of the eigenmodes in the bulk.

We have fitted the radial trajectory of the $\eta$ mesons, as we showed in table \ref{tab:pseudomes} with $\delta_\text{RMS}=\text{21.5}\,\%$. The first excited state, $\eta(550)$, was not well fitted by the model. However, the $\eta'(975)$ meson is well fitted within an error of $1.71\%$. Since our model does not consider any explicit chiral effects via its parameters ($\kappa$, $z_{0}$ and $\Delta_{P}$), we have not addressed chiral anomalies so that the mass difference between $\eta$ and $\eta^{'}$ is to be very well described. 

On the other hand, we have fitted the axial vector meson spectrum. We have included, as a consistency check, other possible $a_1$ candidates. The first state was not well fitted, but this is common in other AdS/QCD soft wall-like models that do not include \emph{vev} or dynamical effects. 

The RMS error associated to this fit is close to 11.2\%. 


For both of these cases, we have taken $I=J=0$ for the $\eta$'s and $I=1,\,J=1$ for the axial particles so that we only have to consider one Regge trajectory for each multiplet. Furthermore, both of these trajectories are linear in the orbital number since the separation between two consecutive poles in the two-point function is almost constant. This is an expected result since both $\eta$ and $a_1$ particles are regular $q\overline{q}$ mesons and these sort of particles are represented through linear Regge trajectories in the $(n,\,M^{2})$ plane \cite{Anisovich:2000kxa}.

After comparing our results with those given in other effective models such as \cite{Parganlija:2012fy}, where twenty-one phenomenological input parameters are taken to properly fit the spectrum due to the huge amount of mesons considered, we infer that the small amount of theoretical parameters taken in our approach must have had an important influence in our results for the masses described in this manuscript. Something similar happens with the radial states in non-conformal models of scalar and vector mesons, where the lowest state of these last ones has an error of approximately 20.5 $\%$ \cite{Cortes:2017lgz}. 

Other non-conformal holographic approaches also take into account more parameters such as quark masses and condensates to fit meson spectra, with both of them coming from an auxiliar scalar field that describes chiral symmetry breaking \cite{Gherghetta:2009ac,Vega:2010ne,Vega:2011tg}. Although having more parameters is very useful to minimize fitting errors, we did not take them into account since our mechanism for symmetry breaking does not have any information about quark structure for the mesons we analyzed.

Finally, we could confirm that chiral symmetry is restored at highly orbital excited states; this is checked after confirming that the degeneracy between very massive chiral partners holds since the squared mass difference between the highest $\rho$ and $a_{1}$ states considered is close to 0.323 GeV$^{2}$, as expected from \cite{Huang:2007fv}. We did not prove this fact with chiral partners such as $f_{0}$'s and $\pi$'s states since we analized the $\eta$ spectrum instead.

\vspace{0.2cm}
\noindent
\textbf{Acknowledgments: }  Authors would like to thank  the financial support given by FONDECYT (Chile) under the Grants No. 1180753 (A. V) and No. 3180592 (M. A. M. C).

\bibliography{biblio.bib}

\end{document}